\documentstyle[aps,twocolumn,prl,floats,epsf]{revtex}
\def\geqap{\,\raise 2pt \hbox{$>\kern-11pt \lower 5pt \hbox{$\sim$}$}\,}
\def\leqap{\,\raise 2pt \hbox{$<\kern-10pt \lower 5pt \hbox{$\sim$}$}\,}
\begin{document}
\draft
\twocolumn[\hsize\textwidth\columnwidth\hsize\csname @twocolumnfalse\endcsname
\title{Theory of Orbital Ordering, Fluctuation and Resonant X-ray Scattering in Manganites}
\author{Sumio Ishihara and Sadamichi Maekawa}
\address{Institute for Materials Research, Tohoku University,  Sendai,
980-8577, Japan}

\date{\today}
\maketitle
\begin{abstract} 
A theory of resonant x-ray scattering in perovskite manganites is developed by 
applying the group theory to the correlation functions of the pseudospin operators for 
the orbital degree of freedom. 
It is shown that static and dynamical informations of the orbital state 
are directly obtained from the elastic, diffuse and inelastic scatterings 
due to the tensor character of the scattering factor. 
We propose that the interaction and its anisotropy between orbitals 
are directly identified by the intensity contour of the diffuse scattering 
in the momentum space. 
\end{abstract}
\pacs{PACS numbers: 75.30.Vn, 71.10.-w, 78.70.Ck} 
]
\narrowtext
Since the discovery of colossal magnetoresistance, 
studies of perovskite manganites have been revived. 
One of the key factors to uncover the 
dramatic and fruitful phenomena in manganites 
is the orbital degree of freedom of manganese ions. 
The collective orbital ordered state as well as its fluctuation 
and excitation are investigated theoretically and experimentally. 
Recently, Murakami {\it et al.} have directly observed the orbital ordering (OO) 
by the resonant x-ray scattering (RXS) \cite{murakami1}. 
This technique reveals several characteristics of the degree of freedom 
through a variety of experiments, i.e.  
the polarization and temperature dependences of the scattering intensity 
and the diffuse scattering \cite{murakami2,endoh,nakamura,zimmermann,wakabayashi}.   
\par
As compared with other scattering experiments, 
RXS is appropriate for observation of the orbital orderings and fluctuations. 
This is owing to the unique characters that (1) an anisotropy of the local electronic structure 
is detectable due to the tensor character of the scattering factor 
together with a short wave length of x-ray \cite{templeton,dmitrienko,ishihara3},  
and (2) this is a site selective technique by tuning 
an energy of the incident x-ray at the Mn $K$-edge. 
However, in spite of recent several theoretical approaches \cite{ishihara3,ishihara4,elfimov,ovchinnikova}, 
a general theory of RXS by the orbital degree of freedom 
has not been presented. 
\par
In this letter, 
we develop a theory of RXS in manganites, for the first time, 
by applying the group theory to the correlation function of the 
pseudospin (PS) operators 
for the orbital degree of freedom. 
The obtained form of the cross section corresponds to those in the neutron scattering  
and in the conventional x-ray/electron scatterings 
as probes to detect spin and charge degrees of freedom, respectively. 
We show that static and dynamical informations of the orbital state are directly 
obtained by the elastic, diffuse and inelastic scatterings.   
\par
Let us formulate the scattering cross section in RXS. 
We consider the scattering of x-ray with momentum $\vec k_i$, energy $\omega_i$
and polarization $\lambda_i$ to $\vec k_f$, $\omega_f$ and $\lambda_f$. 
The electronic state at the initial (final, intermediate) state in the scattering is described 
by $| i\rangle$ ($| f \rangle$, $|m \rangle$) 
with energy $\varepsilon_{i}$ ($\varepsilon_f$, $\varepsilon_m$). 
The scattering cross section is given by \cite{blume,kolpakov}
\begin{equation}
{d^2 \sigma \over d \Omega d \omega_f}= A {\omega_f \over \omega_i}  \sum_{| f \rangle }|S|^2 
 \delta(\varepsilon_f+\omega_f-\varepsilon_i-\omega_i) , 
 \label{eq:sigma}
\end{equation}
where 
\begin{eqnarray}
S&=&  \sum_m \Biggl \{ 
{ \langle f | \vec j_{-k_i} \cdot \vec e_{k_i \lambda_i }|m \rangle 
  \langle m | \vec j_{ k_f} \cdot \vec e_{k_f \lambda_f}  |i \rangle 
  \over \varepsilon_i-\varepsilon_m-\omega_f}
  \nonumber \\ 
&\ &\ \ \  \ \ +
{\langle f | \vec j_{k_f}  \cdot \vec e_{k_f \lambda_f} | m \rangle 
 \langle m | \vec j_{-k_i} \cdot \vec e_{k_i \lambda_i}  | i \rangle 
 \over \varepsilon_i-\varepsilon_m+\omega_i+i \Gamma}
 \Biggr \} ,
 \label{eq:s}  
\end{eqnarray}
and $A=(e^2/mc^2)^2$. 
$\Gamma$ indicates the damping of a core hole, $\vec e_{ k \lambda}$ 
is the polarization vector of x-ray and 
$\vec j_k$ is the current operator divided by a factor $e/\sqrt{m}$.
Eq.~(\ref{eq:sigma}) together with Eq.~(\ref{eq:s}) is rewritten by the correlation function 
of the electronic polarizability \cite{kolpakov,moriya,dmitrienko} as 
\begin{eqnarray}
{d^2 \sigma \over d \Omega d \omega_f}&=&A {\omega_f \over \omega_i}
\sum_{\alpha \beta \alpha' \beta'}
P_{\beta' \alpha' } P_{ \beta \alpha}
\Pi_{\beta' \alpha' \beta \alpha}(\omega, \vec K) , 
\label{eq:sigma2}
\end{eqnarray}
where
\begin{eqnarray}
\Pi_{\beta' \alpha' \beta \alpha}(\omega, \vec K)&=&{1 \over 2 \pi}
\int dt e^{i \omega t} \sum_{ll'} e^{-i\vec K \cdot (\vec r_{l'}-\vec r_{l})}
\nonumber \\
&\times& \langle i |\alpha_{l' \beta' \alpha'}(t)^\dagger 
\alpha_{l \beta \alpha}(0)| i  \rangle , 
\label{eq:pialal}
\end{eqnarray}
$\vec K=\vec k_i-\vec k_f$, $\omega=\omega_i-\omega_f$ 
and $P_{\beta \alpha }=(\vec e_{ k_f \lambda_f})_{\beta}   (\vec e_{k_i \lambda_i})_{\alpha}$. 
The polarizability operator at site $l$ is defined by 
$ \alpha_{l \beta \alpha }(t)=e^{iH_et} \alpha_{l \beta \alpha}^{(\omega_i) } e^{-iH_et}$ 
where $H_e$ is the electronic part of the Hamiltonian and 
\begin{equation}
 \alpha_{l \beta \alpha }^{(\omega_i)}=i \int_{-\infty}^{0} dt  e^{-i \omega_i t} 
 [ j_{l \beta}(0),  j_{l \alpha }(t) ]  , 
\end{equation}
with $[ \cdots ]$ being a commutator. 
Since RXS at the Mn $K$-edge is dominated by the electric dipole transition, 
$\vec j_k$ is given by the local current operator which is independent of $\vec k$ as 
$\vec j_k=\sum_l e^{-i \vec k \cdot \vec r_l}\vec j_l$.
\par
The polarizability is expanded in terms of 
the PS operators for the orbital degree of freedom 
by utilizing the group theoretical analyses \cite{raman}. 
We consider the system where the point symmetry 
around a Mn ion is $O_h$ 
and the doubly degenerate orbitals with the $E_g$ symmetry exist in each Mn$^{3+}$ ion. 
The two bases in the $E_g$ symmetry are denoted by $e_{gu}$ and $e_{gv}$ corresponding  
to the $3d_{3z^2-r^2}$ and $3d_{x^2-y^2}$ orbitals, respectively.  
The PS operators are defined as   
$T_{l \mu}={1 \over 2} \sum_{\gamma \gamma' \sigma}
d_{l \gamma \sigma}^\dagger (\sigma_\mu)_{\gamma \gamma'}
d_{l \gamma' \sigma}$ for $\mu=(0,x,y,z)$ where 
$\sigma_0$ is a unit matrix, $\sigma_{\nu}$ $(\nu=x,y,z)$ 
are the Pauli matrix and $d_{l \gamma \sigma}$ 
is the annihilation operator of the $e_g$ electron 
at site $l$ with orbital $\gamma$ and spin $\sigma$. 
These operators have the $A_{1g}$, $E_{gv}$, $A_{2g}$, and $E_{gu}$ symmetries 
for $\mu=$0, $x$, $y$ and $z$, respectively. 
The tensor part with respect to the polarization of x-ray has 
the $T_{1u} \times T_{1u}$ symmetry which is 
reduced to $A_{1g}+E_g+T_{1g}+T_{2g}$. 
Since  the polarizability should have the $A_{1g}$ symmetry, 
$\alpha_{l \beta \alpha}$ associated with a PS operator 
is represented as 
\begin{eqnarray}
\alpha_{l \beta \alpha }
&=& \delta_{\alpha \beta} I_{A_{1g}} T_{l0} \nonumber \\
&+& \delta_{\alpha \beta} I_{E_g} \Bigl ( \cos{2 \pi n_\alpha \over 3} T_{lz}
                  -\sin{2 \pi n_\alpha  \over 3} T_{lx} \Bigr ) , 
\label{eq:alpht}
\end{eqnarray}
where $(n_x,n_y,n_z)=(1,2,3)$ 
and $I_{A_{1g}(E_g)}$ is a coupling constant. 
We note the following characteristics 
in Eq.~(\ref{eq:alpht}):  
(1) higher order terms with respect to $T_{l \mu}$ at the same site 
are reduced to this form by using the operator algebra, 
(2) $T_{ly}$ does not appear, 
because the tensor part of $\alpha_{l \beta \alpha}$ 
does not include the $A_{2g}$ symmetry, and 
(3) spin operators at a site are not included, 
because the spin-orbit coupling is quenched in a Mn ion. 
We neglect terms including $T_{m \ne l \mu}$ caused by the higher order 
processes of the electron hopping and the Coulomb interactions. 
By using Eq.~(\ref{eq:alpht}),  
the cross section in Eq.~(\ref{eq:sigma2}) is expressed 
by the correlation function of the PS operators as  
\begin{eqnarray}
\Pi_{ \beta' \alpha' \beta \alpha}(\omega, \vec K  )&=& \delta_{\beta' \alpha'} \delta_{\beta \alpha}
{1 \over 2\pi} \int dt e^{i\omega t}
\sum_{l l'}e^{-i \vec K \cdot (\vec r_{l'}-\vec r_{l})} \nonumber \\
& \times &
\sum_{\gamma, \gamma'=0,x,z}
I_{\gamma' \alpha'} I_{\gamma \alpha}  
\langle T_{l' \gamma'}(t) T_{l \gamma}(0)  \rangle , 
\label{eq:pitt}
\end{eqnarray}
with 
$I_{0 \alpha}=I_{A_{1g}}$ and 
$I_{x(z) \alpha}=I_{E_g} \cos(-\sin){2 \pi n_\alpha \over 3}$. 
$\langle i | \cdots | i \rangle$ is replaced by 
the thermal average $\langle \cdots \rangle$. 
When the energy of the scattered x-ray is integrated out, 
the cross section is given by  
\begin{equation}
{d \sigma \over d \Omega} = A\sum_{\alpha \alpha'}
P_{\alpha' \alpha'}
P_{\alpha \alpha}
\sum_{\gamma , \gamma'=0,x,z} I_{\gamma' \alpha'}
I_{\gamma \alpha} S_{\gamma' \gamma}(\vec K) , 
\label{eq:static}
\end{equation}
with $S_{\gamma' \gamma}(\vec K)= \sum_{l l'} 
e^{-i \vec K \cdot( \vec r_{l'}-\vec r_{l})}
\langle T_{l' \gamma'} T_{l \gamma}\rangle$. 
The present results suggest that RXS directly observes the dynamical correlation function  
of the PS operators as a function of $\vec K$ and $\omega$.
This is highly in contrast to the Raman scattering which 
is limited to $\vec K=0$. 
\par
The convincing candidate for the 
microscopic origin of the $T_{l \mu}$ dependent polarizability 
is the Coulomb interaction between Mn $4p$ and 
Mn $3d$ electrons 
$\langle \gamma_2 \alpha_2 | V | \gamma_1 \alpha_1 \rangle$ \cite{ishihara3,ishihara4,ishihara5}
where $\gamma_{1(2)}$ ($\alpha_{1(2)}$) indicates the orbital states of the Mn $3d$ ($4p$) 
electron, and $V$ is the Coulomb interaction. 
The diagonal and off-diagonal components of the interaction are expressed by 
the Slater integral $F_{0(2)}$ between the electrons as  
$\langle \gamma \alpha | V |\gamma \alpha \rangle=F_0+{4 \over 35}F_2 
\cos(\theta_\gamma - {2 \pi m_\alpha \over 3})$  
and 
$\langle {\bar \gamma} \alpha | V |\gamma \alpha \rangle={4 \over 35}F_2 
\sin(\theta_\gamma - {2 \pi m_\alpha \over 3})$, respectively, 
where $(m_x,m_y,m_z)=(1,2,3)$, $\theta_\gamma$ is an angle 
describing the occupied orbital states as 
$|3d_{\gamma} \rangle =\cos({\theta \over 2})|d_{3z^2-r^2}\rangle
-\sin({\theta \over 2})|d_{x^2-y^2}\rangle$, and 
$\gamma$ (${\bar \gamma}$) indicates the occupied (unoccupied) orbital. 
The first and second terms of $\langle \gamma \alpha | V |\gamma \alpha \rangle$ 
and $\langle {\bar \gamma} \alpha | V |\gamma \alpha \rangle$ 
give rise to the $T_{l0}$, $T_{lz}$ and $T_{lx}$ 
dependences of $\alpha_{l \beta \alpha }$, respectively. 
\par
We first apply the scattering cross section to 
the elastic scattering below the OO temperature $T_{OO}$. 
Let us consider the charge and orbital ordered states where 
both the reciprocal lattice vectors in the superlattices $\vec G_c$ and $\vec G_o$  
are assumed to be parallel to the $\vec a+\vec b$ axis in the cubic coordinate. 
The orbital ordered state with two kinds of sublattices are 
denoted as $(\theta_A/\theta_B)$, 
and $\langle {\widetilde T_z} \rangle={1 \over 2} \sum_{l=A,B} (\cos \theta_l \langle T_{l z} \rangle
                                       -\sin \theta_l \langle T_{l x} \rangle) $ 
is adopted as an order parameter. 
In the conventional experimental arrangement in RXS, 
the azimuthal angle $\phi$ is the rotating one of the sample with 
respect to the scattering vector, and the incident x-ray has a $\sigma$ polarization \cite{murakami1,ishihara3}. 
From Eq.~(\ref{eq:static}), the cross section is obtained as 
\begin{eqnarray}
{d \sigma \over d \Omega } \Bigr |_{\sigma \rightarrow \lambda_f}=A N^2 \Bigl |I^c_{\lambda_f} (\vec K)
+I^o_{\lambda_f} (\vec K) \Bigr |^2 . 
\label{eq:cs}
\end{eqnarray}
The charge components of the scattering amplitude are  
$I^c_{\sigma}(\vec G_o)=2I_{A_{1g}} \langle T_0 \rangle$ and 
$I^c_{\pi}(\vec G_o)=I^c_{\sigma,\pi}(\vec G_c)=0$, and 
the orbital ones are  
$I^o_{\lambda_f}(\vec K)=I_{E_{g}} \langle {\widetilde T_{z}} \rangle 
\{ (\cos \theta_A \pm \cos \theta_B)  P^c_{\lambda_f} 
+(\sin \theta_A \pm \sin \theta_B)  P^s_{\lambda_f} \} $.  
Here, $N$ indicates a number of Mn ions in the $A$ sublattice, 
$\langle T_0 \rangle$ is the order parameter for charge, 
$+(-)$ is for $\vec K=\vec G_{c(o)}$ and 
$P^{c(s)}_{\lambda_f}$ is a function of the azimuthal  
and scattering angles \cite{polar}. 
In the following, we consider the cases of $\vec K=\vec G_o$ and $\vec G_c$, respectively. 
(1) $ \vec K=\vec G_o $: 
Through the polarization measurements, 
we can judge whether the $(\theta_A/-\theta_A)$-type 
of the orbital ordered state, 
such as the $(d_{3x^2-r^2}/d_{3y^2-r^2})$- and $(d_{x^2-z^2}/d_{y^2-z^2})$-types,  
is realized or not. 
By inserting the relation $\theta_B=-\theta_A$ into Eq.~(\ref{eq:cs}), 
we derive $d \sigma /d \Omega|_{\sigma \rightarrow \sigma}=0$  
and 
\begin{equation}
{d \sigma  \over d \Omega} \Bigr |_{\sigma \rightarrow \pi}
=A N ^2 3 I_{E_g}^2\langle {\widetilde T_{z}} \rangle^2 \sin^2 
\theta_A \sin^2 \phi \cos^2 \theta_s , 
\label{eq:opi}
\end{equation}
which shows a two-fold symmetry for $\phi$.  
These are clearly distinct from the results with the relation  
$\theta_B \ne -\theta_A$; 
$d \sigma /d \Omega|_{\sigma \rightarrow \sigma}$ is finite 
and 
$d \sigma /d \Omega|_{\sigma \rightarrow \pi}$ has  
a component of a four-fold symmetry for $\phi$. 
It is experimentally observed that   
$d \sigma/ d \Omega |_{\sigma \rightarrow \sigma} =0$ and 
$d \sigma/d \Omega |_{\sigma \rightarrow \pi} \propto  \sin^2 \phi$  
\cite{murakami1,murakami2,endoh,nakamura,zimmermann,wakabayashi}.  
Thus, we judge that 
the $(\theta_A/-\theta_A)$-type of the orbital ordered state 
is realized in these compounds, although the value of $\theta_A$ is not determined 
by measurements at this reflection. 
(2) $\vec K=\vec G_c $:
The $(\theta_A/-\theta_A)$-type of the ordered state is assumed.
At $\lambda_f=\sigma$, where the  RXS experiments for the charge 
ordering are carried out \cite{murakami1,nakamura,zimmermann,wakabayashi}, 
both $I^c_{\sigma}$ and $I^o_{\sigma}$ are finite.   
The latter 
depends on $\phi$ and changes its sign with changing $\theta_A$ 
as
$I^o_{\sigma}=I_{E_g} \langle \widetilde T_z \rangle 
{1 \over 2} \cos \theta_A (1+3\cos 2 \phi)$. 
Thus, the interference term of $I^o_\sigma$ and $I^c_\sigma$ 
is available to determine $\theta_A$ 
through measurements of the $\phi$ dependence of the cross section. 
On the other hand, at $\lambda_f=\pi$, 
the charge component is zero and 
the cross section is given by 
\begin{equation}
{d \sigma \over  d \Omega} \Bigr |_{\sigma \rightarrow \pi}=
A N^2 3I_{E_g}^2\langle {\widetilde T_{z}} 
\rangle^2 \cos^2 \theta_A \sin^2 2\phi \sin^2 \theta_s , 
\label{eq:cpi}
\end{equation}
showing a four-fold symmetry for $\phi$.  
By combining the measurements of $d \sigma /d \Omega|_{\sigma \rightarrow \pi}$ 
at $\vec K=\vec G_o$ (Eq.~(\ref{eq:opi})) and at $\vec K=\vec G_c$ (Eq.~(\ref{eq:cpi})), 
a value of $|\tan \theta_A|$ can be obtained. 
\par
As shown in Eq.~(\ref{eq:opi}), 
$d \sigma /d \Omega|_{\sigma \rightarrow \pi}$ 
at $\vec K=\vec G_o$ is proportional to the factor 
$\sin^2 \theta_A$ as well as $\langle \widetilde T_z \rangle^2$. 
It becomes maximum in the state of 
$\{ {1 \over \sqrt{2}}(d_{3z^2-r^2}-d_{x^2-y^2})/
  {1 \over \sqrt{2}}(d_{3z^2-r^2}+d_{x^2-y^2}) \}$-type  
$(\theta_A=\pi/2)$ 
rather than the $(d_{x^2-z^2}/d_{y^2-z^2})$- ($\theta_A=\pi/3$)
or $(d_{3x^2-r^2}/d_{3y^2-r^2})$-types ($\theta_A=2\pi/3$).
It is experimentally reported that 
the cross section largely changes far below $T_{OO}$ 
where $\langle \widetilde T_z \rangle$ 
is expected to be saturated \cite{murakami1,murakami2,endoh,zimmermann,wakabayashi}. 
A dramatic increase in the cross section 
was observed especially in LaMnO$_3$ at low temperatures \cite{murakami2}. 
We propose that a possible mechanism of the anomalous temperature dependences 
is due to the change of the value of $\theta_A$. 
In particular, the coupling between spin and orbital degrees of freedom 
can change $\theta_A$ 
near the magnetic transition temperature. 
In LaMnO$_3$, 
the increase in the factor $\sin^2 \theta_A$ with decrease in  
$\theta_A$ from $2\pi/3$ 
(the ($d_{3x^2-r^2}/d_{3y^2-r^2}$)-type) 
is supposed to be realized 
near the N$\rm \acute e$el temperature of the A-type antiferromagnetic state 
where the $(d_{x^2-z^2}/d_{y^2-z^2})$-type  
$(\theta_A=\pi/3)$ is stabilized \cite{kugel,maezono}.   
\par
%
%
\begin{figure}
\epsfxsize=0.8\columnwidth
\centerline{\epsffile{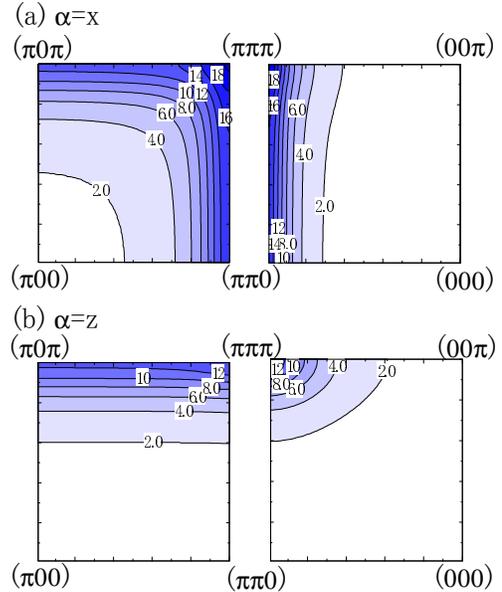}}
\caption{The intensity contour of the diffuse scattering: 
$B_{\alpha \alpha}(\vec K)=(NT \chi_0 I_{E_g}^2)^{-1}\sum_{\gamma' \gamma} 
I_{\gamma' \alpha} I_{\gamma \alpha}S_{\gamma' \gamma}(\vec K)$
 at $T=1.05T_{OO}$. 
(a) $\alpha=x$ and (b) $\alpha=z$ cases. 
}
\end{figure}
We next apply the cross section in Eq.~(\ref{eq:static}) 
to the diffuse scattering.   
Being based on the general model describing the interaction 
between PS operators $H={1 \over 2}\sum_{l' l , \gamma'  \gamma=x,z} T_{l' \gamma'}
J_{\gamma' \gamma} (\vec r_{l'}-\vec r_{l}) T_{l \gamma}$ 
together with the random-phase approximation, 
the static correlation function $S_{\gamma' \gamma}(\vec K)$ is obtained as 
\begin{equation}
S_{\gamma' \gamma}(\vec K)=N\chi_0 T [1+\chi_0 J(\vec K) ]^{-1}_{\gamma' \gamma} , 
\label{eq:sdiffuse}
\end{equation} 
with $\chi_0=1/(4T)$. 
$S_{\gamma' \gamma}(\vec K)$ around $\vec K=\vec G_o$ 
near $T_{OO}$ contributes to the 
diffuse scattering in RXS. 
It is worth to note that, as shown in Eq.~(\ref{eq:static}),
an intensity contour of the diffuse scattering  
directly reflects the momentum dependence of $S_{\gamma' \gamma}(\vec K)$ 
and each element of $S_{\gamma' \gamma}(\vec K)$ with respect to $\gamma$ and $\gamma'$ 
is obtained 
by changing a polarization of x-ray. 
This is owing to the tensor character of the scattering factor 
and is highly in contrast to the conventional x-ray and neutron scatterings \cite{kataoka}; 
Unlike the present case, 
the intensity contour in those cases 
is not directly determined by $S_{\gamma' \gamma}(\vec K)$ 
but the correlation function of charge, 
which connects with $S_{\gamma' \gamma}(\vec K)$ through the 
momentum-dependent coupling constant between 
PS and lattice vibration, and the factor 
$\vec K \cdot \vec e_{jKs}$, where $\vec e_{jKs}$ is the displacement vector 
of the $j$th ion for a normal coordinate with momentum $\vec K$ and mode $s$.  
\par
The above form of the diffuse scattering is applied to a case 
of OO in perovskite manganites where 
the interaction between orbitals is dominantly caused by the electronic process.  
The following Hamiltonian is adopted to calculate the intensity contour microscopically \cite{ishihara1,exchange,endoh};   
\begin{eqnarray}
H=\sum_{\langle ij \rangle } 
\Bigl ( {3 \over 2} J_1^l  
-{1 \over 2} J_2^l \Bigr )
 \tau_i^l \tau_j^l 
 \ , 
\label{eq:hj}
\end{eqnarray}
where $\tau_i^{l}=\cos(2\pi n_l/3)T_{iz}-\sin(2 \pi n_l/3)T_{ix}$ 
with $(n_x, n_y,n_z)=(1,2,3)$,  $l$ indicates a direction of a bond between site $i$ and 
its nearest neighboring $j$, and  
$J_1^l$ and $J_2^l$ are the superexchange-type interactions 
between PS's. 
This model is derived by the perturbational calculation of the electron hopping 
between Mn ions under the strong Coulomb interactions \cite{ishihara1}. 
In Fig.~1, 
we present 
the calculated intensity contour of 
$B_{\alpha \alpha}(\vec K)=(NT \chi_0 I_{E_g}^2)^{-1}\sum_{\gamma' \gamma} 
I_{\gamma' \alpha} I_{\gamma \alpha}S_{\gamma' \gamma}(\vec K)$ 
at $T=1.05T_{OO}$ by which 
the cross section is represented as   
$d\sigma/d \Omega=A\sum_{\alpha' \alpha}P_{\alpha' \alpha'}P_{\alpha \alpha}
NT\chi_0I_{E_g}^2B_{\alpha' \alpha}(\vec K)$. 
OO occurs at $\vec G_o=(\pi \pi \pi)$ below $T_{OO}$. 
Strong intensity appears 
along the $(\pi \pi \pi)-(\pi \pi 0)$ direction and other two equivalent ones. 
This feature is attributed to the unique character of the 
superexchange-type interaction between PS's 
that $J_{xx}(\vec r_l-\vec r_{l'}=a \hat z)=0$, 
that is, the hopping integral between the $d_{x^2-y^2}$ 
and $d_{x^2-y^2(3z^2-r^2)}$ orbitals is zero in the $z$ direction \cite{ishihara2,diffuse}. 
The diffuse scattering in RXS has been experimentally observed in Pr$_{1-x}$Ca$_x$MnO$_3$ \cite{zimmermann},  
although the measurements were carried out along 
one direction in the Brillouin zone.
In this technique,
the critical orbital fluctuation is directly observed around principal $\vec G_o$'s  
where the conventional x-ray scattering is forbidden in perovskite crystals. 
It is proposed that 
through observations of the characteristic features along    
the $(\pi \pi \pi)-(\pi \pi 0)$ and other equivalent directions with changing 
a polarization of x-ray, 
we can identify the superexchange-type interaction between orbitals. 
\par
The present form of the cross section 
is also appropriate for the study of the inelastic RXS due to  
the collective excitation in the orbital 
ordered state termed orbital wave (OW). 
The dispersion relations of OW have been  
examined in the Hamiltonian Eq.~(\ref{eq:hj}) with 
spin degree of freedom \cite{ishihara5}. 
By applying the Holstein-Primakoff transformation to the PS operator in Eq.~(\ref{eq:pitt}), 
the scattering cross section for OW is given by 
Eq.~(\ref{eq:sigma2}) with 
\begin{eqnarray}
\Pi_{\alpha' \alpha}(\omega, \vec K)&=&{N \over 4} I_{E_g}^2 \sum_{\vec q  \vec G} 
 \sum_{\mu \mu'} e^{-i\vec G \cdot (\vec r_{\mu'} -\vec r_{\mu})}
 S_{\alpha' {\mu'}} S_{\alpha {\mu}}
\nonumber \\ 
&  \times & \sum_\nu \Bigl \{  X_{\mu' \nu} X_{\mu \nu}^\ast  
(n_{\vec q}^\nu+1) \delta(\omega-\omega_{\vec q}^\nu) \delta_{\vec K-\vec q+\vec G}
\nonumber \\
& \ &  \ \ \ \ \  +X_{\mu' \nu}^\ast X_{\mu \nu} n_{\vec q}^\nu 
\delta(\omega+\omega_{\vec q}^\nu) \delta_{\vec K+\vec q+\vec G})
\Bigr \} , 
\label{eq:inela}
\end{eqnarray}
where $\mu $ ($\nu$) indicates the orbital sublattice (the mode of OW),  
$\omega_{\vec q}^\nu$ ($n_{\vec q}^\nu$) 
is the energy (number) of OW 
and $S_{\alpha \mu}=\sin(2\pi n_\alpha/3-\theta_\mu)$. 
$X_{\mu \nu}=V_{\mu \nu}+W_{\mu \nu}^\ast$ are the coefficients in 
the Bogoliubov transformation defined as 
$a_{\vec q}^\mu =\sum_\nu (V_{\mu \nu} \alpha_{\vec q}^\nu+W_{\mu \nu} \alpha_{-\vec q}^{\nu \dagger})$ 
where $a_{\vec q}^\mu$ ($\alpha_{\vec q}^\nu$) is  
the annihilation operator of the Holstein-Primakoff 
boson (the orbital wave). 
The above form of the cross section is analogous to 
that in the one-magnon neutron scattering 
unlike in the two-magnon Raman one, 
since one OW is possible to be excited in RXS. 
However, there is a clear distinction from the neutron scattering;   
the $y$ component of the operator $T_y$ is not included. 
This characteristic brings about the selection rule  
between $\vec K$ and the modes of OW wave as follows. 
We consider the orbital ordered state of the  
$(\theta_A/\theta_A+\pi)$-type with $\vec G_o=(\pi \pi \pi)$ 
theoretically predicted in the ferromagnetic insulating state. 
It is shown from Eq.~(\ref{eq:inela}) that the acoustic (optical) 
mode of OW  
is separately measured at $h+k+l$=odd (even). 
This is because $T_{ly}$ does not appear in the cross section.  
\par
In summary, we develop a theory of RXS in perovskite manganites, 
for the first time, by applying the group theory to the 
correlation function of the PS operators. 
The tensor character of the scattering factor is essential 
in several kinds of scatterings due to the orbital degree of freedom. 
The present theory is applied easily to RXS in other compounds with the orbital degree of freedom. 
\par
The authors thank Y.~Murakami, Y.~Endoh, T.~Arima, K.~Hirota, M.~Blume, 
D.~Gibbs and J.~P.~Hill for their valuable discussions. 
The work was supported by Grant-in-Aid for Scientific Research Priority Area from the 
Ministry of Education, Science and Culture of Japan, CREST and NEDO. 
Part of the numerical calculation was performed in the HITACS-3800/380 
superconputing facilities in IMR, Tohoku Univ.
\narrowtext

\begin{references}
\bibitem{polar}
$P^c_\sigma=-{1 \over 2} \sin^2 \phi+\cos^2 \phi$, 
$P^c_\pi=-{3 \over 2 }\sin \phi \cos \phi \sin \theta_s$, 
$P^s_\sigma    =0$ and 
$P^s_\pi    =-{\sqrt{3} \over 2 }\sin \phi \cos \theta_s$  
where $\phi$ and $\theta_s$ are the azimuthal and scattering angles, 
respectively.
\bibitem{exchange}
$J_1^l=t_{0l}^2/(U'-I)$ and $J_2^l=t_{0l}^2/U$ 
where $t_{0l}$ is the hopping integral between the $3l^2-r^2$ orbitals
in direction $l$ and $U$, $U'$ and $I$ are the intra-site Coulomb interactions 
in a Mn ion. 
%
The spin correlation in the Hamiltonian presented in Ref.~\onlinecite{ishihara1}
$\vec S_i \cdot \vec S_j$ is replaced by zero, 
since OO appears in the paramagnetic phase in several insulating manganites.
A weak anisotropy in the exchange interactions is introduced 
by $J_{1}^{x(y)}/J_{1}^z=J_{2}^{x(y)}/J_{2}^z=\delta_{x(y)}$ 
to remove degeneracy of the orbital ordered state. 
These values are chosen to be $\delta_x=1.05$ and $\delta_y=1.025$ in Fig.~1.
\bibitem{diffuse}
In the actual compounds, 
the tilting of a MnO$_6$ octahedron may cause a small finite value of 
$J_{xx}(a \hat z)$.
\end{references}
\end{document}